\documentclass[aps,prd,nofootinbib,showpacs,showkeys,preprintnumbers,amsmath,amssymb,superscriptaddress]{revtex4-1}

\usepackage{url}
\usepackage{amsmath,amsfonts,amssymb,amsthm}
\usepackage{bm,bbm,bbold,braket,xcolor,mathrsfs}
\usepackage{graphicx}
\usepackage{array}
\usepackage{tikz}
\usepackage{float}
\usepackage{color}

\usetikzlibrary{positioning,arrows}
\usetikzlibrary{decorations.pathmorphing}
\usetikzlibrary{decorations.markings}
\tikzset{
  particle/.style={thick,draw=black,
    postaction={decorate},
    decoration={markings,mark=at position .5 with
      {\arrow[black]{triangle 45}}}},
  photon/.style={decorate,
    draw=black,
    decoration={coil,aspect=0}}
}

\tikzstyle directed=[postaction={decorate,decoration={markings,
    mark=at position .65 with {\arrow[arrowstyle]{stealth}}}}]
    
\tikzstyle reverse directed=[postaction={decorate,decoration={markings,
    mark=at position .65 with {\arrowreversed[arrowstyle]{stealth};}}}]
    
\usetikzlibrary{arrows,shapes}
\usetikzlibrary{trees}
\usetikzlibrary{matrix,arrows} 				
\usetikzlibrary{positioning}				
\usetikzlibrary{calc,through}				
\usetikzlibrary{decorations.pathreplacing}  
\usepackage{pgffor}							
\usetikzlibrary{decorations.markings}   
\tikzset{
    vector/.style={decorate, decoration={snake}, draw},
	provector/.style={decorate, decoration={snake,amplitude=2.5pt}, draw},
	antivector/.style={decorate, decoration={snake,amplitude=-2.5pt}, draw},
    fermion/.style={draw=black, postaction={decorate},
        decoration={markings,mark=at position .55 with {\arrow[draw=black]{>}}}},
    fermionbar/.style={draw=black, postaction={decorate},
        decoration={markings,mark=at position .55 with {\arrow[draw=black]{<}}}},
    fermionnoarrow/.style={draw=black},
    gluon/.style={decorate, draw=black,
        decoration={coil,amplitude=4pt, segment length=5pt}},
    scalar/.style={dashed,draw=black, postaction={decorate},
        decoration={markings,mark=at position .55 with {\arrow[draw=black]{>}}}},
    scalarbar/.style={dashed,draw=black, postaction={decorate},
        decoration={markings,mark=at position .55 with {\arrow[draw=black]{<}}}},
    scalarnoarrow/.style={dashed,draw=black},
    electron/.style={draw=black, postaction={decorate},
        decoration={markings,mark=at position .55 with {\arrow[draw=black]{>}}}},
	bigvector/.style={decorate, decoration={snake,amplitude=4pt}, draw},
}

\usepackage{siunitx}
\usepackage{hyperref}
\hypersetup{linktocpage,colorlinks=true,allcolors=blue}

\newcommand{\Lag}{\mathscr{L}}
\DeclareMathOperator{\Tr}{Tr}

\DeclareSIUnit\parsec{pc}
\DeclareSIUnit\Mpc{\mega\parsec}
\DeclareSIUnit\lightyear{ly}

\begin{document}
\title{Limits on beyond standard model messengers as ultra high energy cosmic rays}

\author{Oscar \surname{Castillo-Felisola}}
\affiliation{Departamento de F\'isica, Universidad T\'ecnica Federico
  Santa Mar\'ia, Casilla 110-V, Valpara\'iso, Chile}
\affiliation{Centro Cient\'ifico Tecnol\'ogico de Valpara\'iso,
  \\ Casilla 110-V, Valpara\'iso, Chile}
\email{o.castillo.felisola@gmail.com}
\author{Crist\'obal \surname{Corral}}
\affiliation{Instituto de Ciencias Nucleares, Universidad Nacional
  Aut\'onoma de M\'exico, \\ Apartado Postal 70-543, Ciudad de
  M\'exico, 04510, M\'exico}
\email{cristobal.corral@correo.nucleares.unam.mx}
\author{Piotr \surname{Homola}}
\affiliation{Institute of Nuclear Physics, Polish Academy of Sciences,
  Poland}
\email{piotr.homola@ifj.edu.pl}
\author{Jilberto \surname{Zamora-Saa}}
\affiliation{Dzhelepov Laboratory of Nuclear Problems, Joint Institute
  for Nuclear Research, Dubna, Russia}
\email{jzamorasaa@jinr.ru}

\begin{abstract}
Up to date, there is no consensus regarding the origin of ultra high energy cosmic rays (UHECR) beyond the Greisen--Zatsepin--Kuzmin (GZK) limit. In order for these UHECR to reach the Earth, an extremely suppressed interaction between them and the cosmic microwave background (CMB) is required, which is impossible for standard model (SM) particles, except neutrinos. In this letter, we present constraints on the parameter space of models involving axion-like particles and dark photons, as candidates for UHECRs, by assuming that these particles can traverse the visible Universe without decaying. In the case of axion-like particles, the constraints are tighter than those given in previous works. We present for the first time constraints on the parameter space of dark photon models, by considering their kinetic term mixing with SM photons.
\end{abstract}

\pacs{}
\keywords{Axion-like particles, Dark Photons, Cosmic Rays}

\maketitle

\section{Introduction\label{sec.intro}}

According to the standard model of particle physics, if cosmic rays
are primarily composed of protons, there should be a bound on the
maximum energy of the cosmic rays coming from distances greater than
\SI{50}{\Mpc}. This bound was first established in
Refs.~\cite{Greisen:1966jv,Zatsepin:1966jv}, and a more accurate estimation
was reported later in~\cite{Stecker:1968uc} giving \SI{100}{\Mpc}, and it is known as the
GZK limit. In spite of this, UHECR with energies above the GZK limit have been
detected from places where apparently there are no nearby
sources~\cite{Takeda:1998ps,Zotov2017,Verzi:2017hro}. It seems
therefore, that there is a missing piece in our understanding of the
sourcing, nature, and/or propagation of the cosmic rays.

If UHECR events present a small-scale clustering, their sources
could be considered as point-like at cosmological
scales~\cite{Tinyakov:2001ic,Abreu:2012zza,Tkachev:2011fla,AbuZayyad:2012hv},
and it has been suggested---based on coincidence of arrival
direction---that certain astrophysical object could act as sources of some of
the highest energy
events~\cite{Elbert:1994zv,Albuquerque:2010rq,Anchordoqui:2007tn,Ryabov:2006pk}. Nonetheless,
these sources are at red-shift \(z>0.1\),\footnote{Equivalent to a
  comoving radial distance of \SI{421.3}{\Mpc}.} exceeding the GZK
horizon (\(R_{\textsc{gzk}} \approx \SI{100}{\Mpc}\)). Indeed, the Pierre Auger Collaboration 
has recently reported an anisotropy in the arrival directions of the UHECRs with more 
than $5.2\sigma$ of significance~\cite{Aab:2017tyv}, supporting the hypothesis 
of their extragalactic origin. This would mean that it is difficult for
primary ultra high energy particles to be
protons~\cite{Gelmini:2007jy}, since for energies around
\SI{e20}{\eV}, the attenuation length is approximately
\(R_{\textsc{gzk}}\) due to the interaction with extragalactic radio
background. On the other hand, the radio background can be simulated
by means of numerical propagation codes~\cite{Kalashev:2000af,Aloisio:2012wj}, and
they show that it is very unlikely for UHECR to be photons~\cite{Aab:2016agp}.

Within the SM, the only particle that can reach our galaxy
without (significant) loss of energy are neutrinos. Therefore, 
different scenarios modeling UHECR by neutrinos have been proposed. In
one of them, neutrinos produce nucleons and photons via resonant
\(Z\)-production with relic neutrinos clustered within approximately
\(\SI{50}{\Mpc}\) from the Earth, giving rise to angular correlations
with high redshift
sources~\cite{Weiler:1982qy,Weiler:1983xx,Aloisio:2015ega}. However,
in order for this model to be compatible with experimental data, 
a huge neutrino flux should be produced at the source, along with a 
clustering of relic neutrinos~\cite{Aloisio:2016hng,Gelmini:2007jy,BlancoPillado:1999yb}. Other models consider extradimensional scenarios, where an enhancement 
of high-energy neutrino-nucleon cross sections can be produced by the exchange of
Kaluza--Klein graviton modes~\cite{Tyler:2000gt}, or by an exponential increase of the
number of degrees of freedom in string theory
models~\cite{Domokos:1998ry}. Furthermore, if Kaluza--Klein axions are 
considered, their oscillation into photons allows the latter to travel 
large distances without interacting with the CMB, producing in principle
the UHECRs events above the GZK limit~\cite{Nicolaidis:2009zg}. A third 
scenario allowing to circumvent this limit is the Lorentz
invariance violation~\cite{Coleman:1997xq,Bhattacharjee:1998qc,Bietenholz:2008ni,Scully:2008jp,Gorham:2012qs,Rubtsov:2013wwa,Tasson:2014dfa,Rubtsov:2016bea,Mohanmurthy:2016ven}, which is already constrained by astrophysical experiments (see e.g. \cite{Klinkhamer:2008ky}) and can be further tested with global cosmic-ray analyses, as proposed by Cosmic Rays Extremely Distributed Observatory (CREDO) Collaboration \cite{Sushchov:2017kdi,Dhital:2017yuu,Homola:2017pc}, e.g. by considering cascades produced by primary ultra high energy photons in the photon decay scenario (see \cite{Dhital:2017yuu} and references therein). It is worthwhile to note that the latter model implies non-observation of single UHE photons on Earth, as their lifetimes would be extremely short, of the order of 1 second (see e.g. \cite{Jacobson:2005bg} for a review]). 

Another interesting way around to the GZK cutoff, is the possibility that 
UHECR are composed mainly of particles beyond the standard model. In order for these particles to
be acceptable candidates of UHECR, they should: (i) to be stable
enough to reach the Earth from cosmological distances; (ii) interact
very weakly with the CMB and extra
galactic magnetic fields, so that they do not lose much energy; (iii)
be produced with a significant flux at the source; and (iv) interact
sufficiently strong in the near galaxy, with the Sun or Earth magnetic field, 
or with the Earth atmosphere.

Axions, for instance, could be considered as candidates to avoid the
GZK cutoff \cite{Mirizzi:2006zy,Csaki:2003ef}. Nevertheless, it has been shown that it is unlikely that the
axion production, together with their convertion to photons by the
galactic magnetic field, accounts for UHECR within the present
exclusion
limits~\cite{gorbunov01_axion_partic_as_ultrah_energ_cosmic_rays}. Another
possibility is to consider particles with similar features of axions,
such as axion-like particles. This case has been studied in
Refs.~\cite{gorbunov01_axion_partic_as_ultrah_energ_cosmic_rays,Csaki:2003ef,Fairbairn:2009zi},
constraining their parameter space according to current experimental
data.

The aim of this paper is to consider two beyond the standard model particles, such as
axion-like particles and dark photons, as candidates for UHECRs by considering 
that they can traverse the visible Universe without decaying, producing the
detected events on Earth. We present an update on the parameter space
of axion-like particles in light of the new experimental 
results~\cite{Anastassopoulos:2017ftl,Cadamuro:2011fd,DiLuzio:2016sbl} and, 
for the first time, we present constraints on the parameter space of dark photons
characterized by their mass and mixing with ordinary photons.

The paper is structured as follows: In Sec.~\ref{DMDP}, we review a
model of dark photons acting as source
of UHECR. In Sec.~\ref{sec.review_ALP}, we review a few models of
axions and axion-like particles. Next, in Sec.~\ref{astro_impl} we
analyse the astrophysical implications of considering the mentioned
particles as sources of UHECR, constraining the values of the coupling
constants. Finally, we present our concluding remarks in
Sec.~\ref{discussion}.

\section{Dark Photons\label{DMDP}}

The astronomical evidence of matter distribution in
galactic rotation curves, mass-to-light ratios due to gravitational
lensing, measurements of the CMB anisotropy, among others,
suggest the existence of either non-standard matter fields, dubbed
dark matter~\cite{Freese:2008cz}, modified gravitational dynamics~\cite{Famaey:2011kh}, 
or both. Due to the lack of direct evidence, it have not been possible to 
discriminate between the responsibles for these phenomena. In this section, we shall
consider the case where dark matter might explain the aforementioned 
experimental data.

A way to understand the origin of dark matter, is provided by
theoretical models which introduce the concept of \emph{dark} sector,
consisting of singlet fields under the standard model gauge group,
but transforming nontrivially under a \emph{dark} gauge group. In general, 
there is no constraint on the size of this dark sector in
comparison with the visible one. Then, the experimental 
sensitivity depends on the coupling and mass scale. 

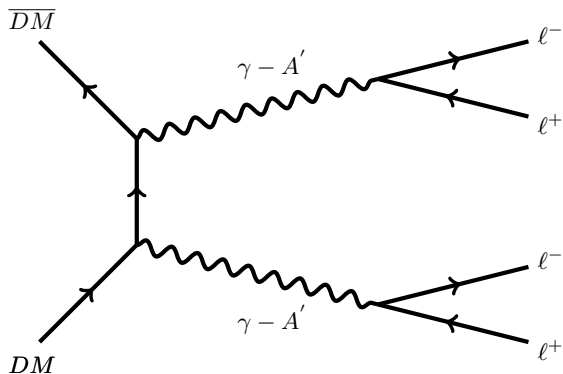
\begin{figure}[t] 
  \centering
  \begin{tikzpicture}[thick,scale=1.0]
    \coordinate (O) at (0,0) ;
   \draw [fermion,ultra thick] (-2.0,-2.0) -- (-0.707,-0.707);
   \node[] at (-2.1,-2.3)  {$DM$};
   \draw [fermion,ultra thick] (-0.707,-0.707) -- (-0.707,0.707);
   \node[] at (-2.1,-2.3)  {$DM$};
   \draw [fermion,ultra thick] (-0.707,0.707)--(-2.0,2,0);
   \node[] at (-2.1,2.3)  {$\overline{DM}$};
   \draw[vector,ultra thick] (-0.707,0.707)--(2.5,1.5);
   \node[] at (1.1,1.7)  {$\gamma-A^{'}$};
    \draw[vector,ultra thick] (-0.707,-0.707)--(2.5,-1.5);
    \node[] at (1.1,-1.7)  {$\gamma-A^{'}$};
    \draw[fermion,ultra thick] (2.5,1.5)--(4.5,2.0);
   \node[] at (4.8,2.1)  {$\ell^-$};
   \draw[fermion,ultra thick] (4.5,1.0)--(2.5,1.5);
   \node[] at (4.8,0.9)  {$\ell^+$};  
   \draw[fermion,ultra thick] (4.5,-2.0)--(2.5,-1.5);
   \node[] at (4.8,-2.1)  {$\ell^+$};
   \draw[fermion,ultra thick] (2.5,-1.5)--(4.5,-1.0);
   \node[] at (4.8,-0.9)  {$\ell^-$};   
\end{tikzpicture}
 \caption{Schematic representation of dark matter annihilation.}
  \label{DP}
\end{figure}

In order to explain the absence of these ultra high energy particles
within the standard model, we focus on a model including an additional $U(1)$ gauge group~\cite{Ackerman:mha}, whose corresponding gauge
boson can aquire its mass through either St\"uckelberg or Higgs mechanism. 
It is assumed that the interaction between
standard model particles and dark matter will be mediated solely by a
new Abelian $U'(1)$ gauge boson, $A^\prime$, dubbed \emph{dark photon},
mixed with ordinary photons $\gamma$. The interaction between $\gamma$ and $A^\prime$
is given by the kinetic mixing~\cite{Holdom:1985ag,Holdom:1986eq}
\begin{equation}
  \label{Lint}
  \mathcal{L}_{\textsc{int}}=-\frac{1}{2} g_{\textsc{dp}} F_{\mu \nu} A^{\prime \mu \nu},
\end{equation}
where $F^{\mu \nu}$ and $A^{\prime \mu \nu}$ are field strength of the ordinary and
dark photon, respectively, and $g_{\textsc{dp}}$ is their mixing
parameter. It is important to note that $g_{\textsc{dp}}$ could be
energy-scale dependent (see Ref.~\cite{Brahmachari:2014aya} and
references therein). The kinetic mixing~\eqref{Lint} induce a $\gamma - A^\prime$ oscillation, similar to the case
of massive neutrinos~\cite{Brahmachari:2014aya}. Although dark photon interactions
have not been detected in experimental searches~\cite{Bjorken:2009,Brown:2015bot}, it does not exclude
that---provided their existence---they could be relevant in higher energy
scales. Thus, any source of photons could produce a kinematically
allowed massive $A^\prime$ state, in accordance with the mixings. Within the
heavy dark matter frameworks, processes as shown in
Fig.~\ref{DP} (dark matter annihilation) can produce UHECR which
energies reaching the scale of heavy dark matter masses
$M_{\textsc{dm}}$, in the range between
\SIrange{e2}{e19}{\GeV}, depending on the dark matter model (see for
example Refs.~\cite{PhysRevLett.116.101302,Peter:2012rz}).

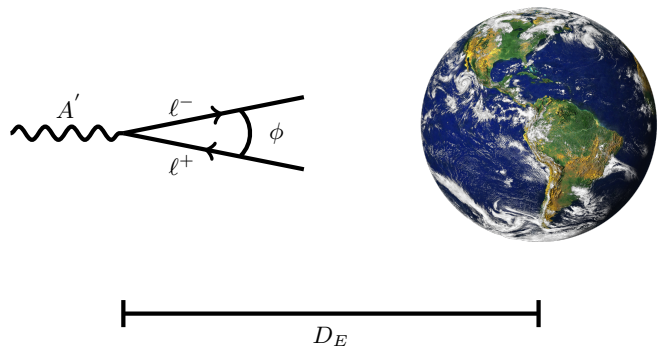
\begin{figure}[t] 
  \centering
\begin{tikzpicture}[thick,scale=0.6]
    \node[anchor=south west,inner sep=0] at (9,0.5) {\includegraphics[width=0.18\textwidth]{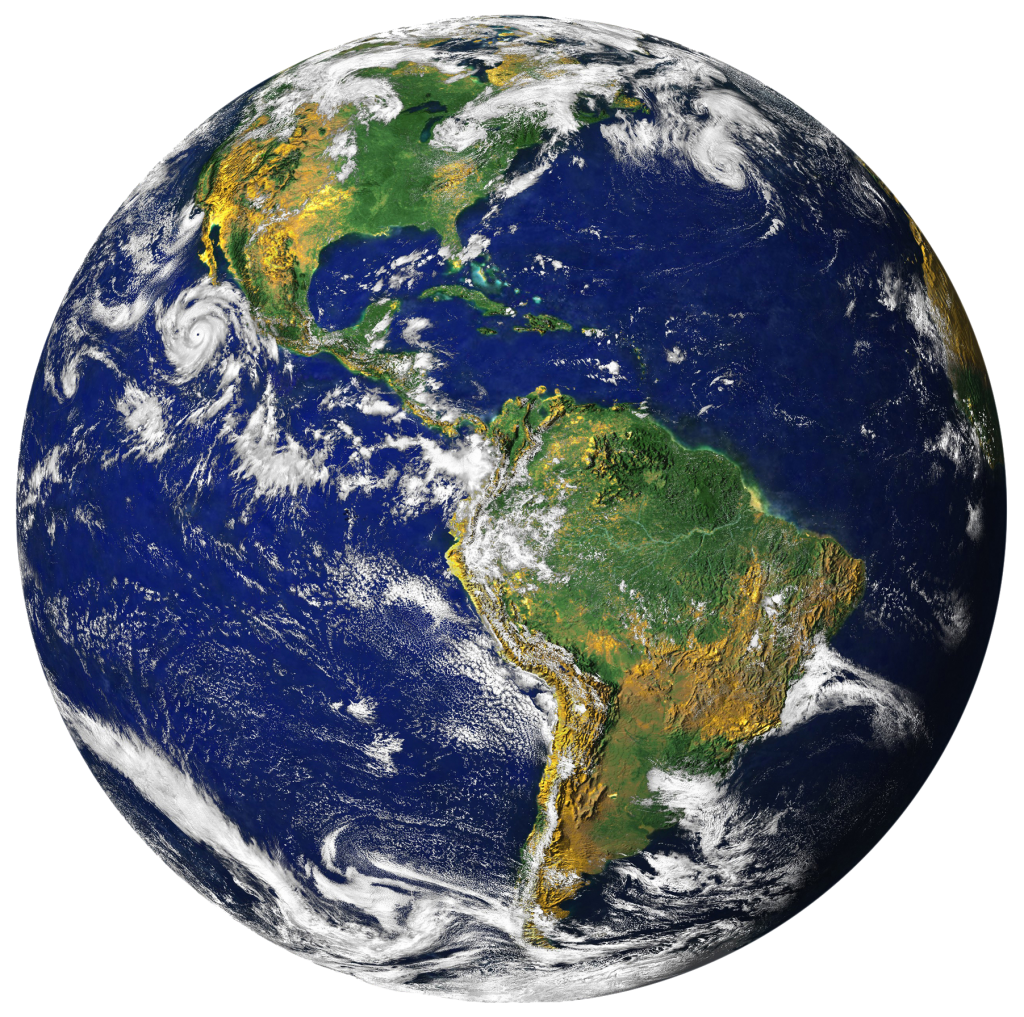}};
    \coordinate (O) at (0,0) ;
   \draw[ultra thick,vector] (0.0,3.0)--(2.5,3.0);
   \node[ultra thick] at (1.3,3.6)  {$A^{'}$};
   \draw[fermion,ultra thick] (2.5,3.0)--(6.5,3.8);
   \node[ultra thick] at (3.8,3.6)  {$\ell^-$};
   \draw[fermion,ultra thick] (6.5,2.2)--(2.5,3.0);
   \node[ultra thick] at (3.8,2.3)  {$\ell^+$};  
   \draw [ultra thick,black] (5.1,3.5) to[out=-45,in=45] (5.1,2.5) ;
   \node[right,black,ultra thick] at (5.5,3.0)  {$\phi$}; 
   \draw [ultra thick,black] (2.5,-1.0) to (11.7,-1.0) ;  
   \draw [ultra thick,black] (2.5,-1.3) to (2.5,-0.7) ;
   \draw [ultra thick,black] (11.7,-1.3) to (11.7,-0.7) ;   
   \node[black,ultra thick] at (7.1,-1.5)  {$D_E$};
\end{tikzpicture}
 \caption{Schematic representation of the decay angle $\phi$. Here, $D_E$ stands for the distance from the dark photon decay vertex to center of the Earth.}
  \label{DPdecay}
\end{figure}

Assuming the existence of dark photons, these oscillate into photons
(and vice-versa) with an efficiency driven by their mass $m_{A^\prime}$ and
the mixing coupling $g_{\textsc{dp}}$. Therefore, dark photons can
decay into secondary lepton pairs through \(A^\prime  \to \ell^+ +\ell^-\), 
providing a way to compare with the collected data of future experiments. 
Thus, it would be interesting to analyse the correlations between anisotropy of 
UHECR sources and regions with high dark matter densities. If such correlation 
is found, it would support the model of dark photons.

The partial decay width of the dark photon (with mass $m_{ A^\prime} \geq 2 m_{\ell} $) into a secondary lepton pair is given
by~\cite{Echenard:2014lma}
\begin{equation}
  \label{DPsDW}
  \Gamma(A^\prime \to \ell^{+} \ell^{-}) = \tfrac{\alpha_{\text{em}}}{3}  m_{ A^\prime}
  g_{\textsc{dp}}^2 \sqrt{1-\tfrac{4 m_{\ell}^2}{m_{A^{'}}^2}} \left(  1
  + \tfrac{2 m_{\ell}^2}{m_{A^{'}}^2}  \right)
\end{equation}
where $m_{\ell}$ is the lepton mass and $\alpha_{\text{em}}$ is the
fine-structure constant of quantum electrodynamics. 

Assuming the dark matter mass to be much larger than dark photon one, implies
the latter to be highly boosted after the dark matter annihilation. Thus, the 
decay angle $\phi$, schematically represented in Fig.~\ref{DPdecay}, is given 
by~\cite{Feng:2015hja}
\begin{align}
\label{decayangle}
 \phi \approx \frac{\sqrt{m_{A'}^2 - 4m_\ell^2}}{m_{\textsc{dm}}},
\end{align}
where $m_{\textsc{dm}}$ is the dark matter mass, the assumption \mbox{$m_{\textsc{dm}}\ggg  m_{ A^\prime}\geq 2 m_{\ell}$
has been considered}, and from now on we use that $m_{\ell} = m_e \approx 0.511$ MeV.


\section{Axions and Axion-like particles\label{sec.review_ALP}}

The quark sector of the standard model possesses two different sources
of charge-parity (\(CP\)) symmetry violation: (i) the
Cabibbo--Kobayashi--Maskawa matrix \(M\), arising from the electroweak
symmetry breaking; and (ii) the nontrivial structure of the quantum
chromodynamics (QCD) vacuum, known as the
\(\theta\)-vacuum~\cite{tHooft76_comput_quant_effec_due_to,tHooft78_errat}. The
Cabibbo--Kobayashi--Maskawa matrix is generically endowed with a complex
phase, and their diagonalization induces a redefined parameter,
\(\bar{\theta} = \theta + \arg\det M\), which encodes these two contributions of
\(CP\) violation within the quark sector. The effective Lagrangian which
accounts for these phenomena is given by
\begin{equation}
  \label{eq.qcd_eff}
  \Lag_{\text{eff}} = \Lag_{\textrm{QCD}} + \bar{\theta} \frac{\alpha_s}{4\pi}\Tr\left[G_{\mu\nu}\tilde{G}^{\mu\nu}\right],
\end{equation}
where \(\tilde{G}^{\mu\nu} \equiv \tfrac{1}{2}\epsilon^{\mu\nu\rho\sigma}G_{\rho\sigma}\), \(\alpha_s =
g^2_s/4\pi\) is the coupling constant of QCD, and the trace is taken
over group indices. The last term in Eq.~\eqref{eq.qcd_eff} is the
Pontryagin density for the \(SU(3)\) group, and it is known as the
\(\bar{\theta}\)-term of QCD. This term is a topological invariant which
can be locally written as a boundary term, adding no dynamics to the
field equations. At the quantum level, however, it contributes to
\(CP\)-odd observables such as the neutron electric dipole
moment~\cite{crewther79_chiral_estim_elect_dipol_momen,crewther80_chiral_estim_elect_dipol_momen}. Its
experimental value has been highly constrained, giving an upper limit
of
\(|\bar{\theta}|<10^{-10}\)~\cite{baker06_improv_exper_limit_elect_dipol_momen_neutr}. The
strong \(CP\) problem is known as the lack of explanation for the
smallness of \(\bar{\theta}\), in order to fit its experimental value.

One of the most popular solutions to this problem was proposed by Peccei and
Quinn, by introducing an additional global axial symmetry to the
standard model Lagrangian~\cite{Peccei:1977hh,Peccei:1977ur}. If this
symmetry were exact, one would be able to rotate the
\(\bar{\theta}\)-parameter away. However, it is clear that this symmetry cannot remain
unbroken. Peccei and Quinn showed that one is still able to rotate the
\(\bar{\theta}\)-term away if such a symmetry is spontaneously broken,
while the (pseudo-)Nambu--Goldstone boson associated with the breakdown
of such a symmetry, the axion~\cite{Weinberg:1977ma,Wilczek:1977pj},
replaces the \(\bar{\theta}\)-parameter by a dynamical field, i.e.,
$\bar{\theta} \to \tfrac{a(x)}{f_a}$,
where \(f_a\) is the scale of the axial symmetry
breaking. Nonperturbative effects of QCD generate a potential for the
axion, which selects a vacuum expectation value that cancels the \(\bar{\theta}\)-angle exactly,
solving the strong CP problem dynamically.\footnote{This potential also generates a mass for the axion, determined solely by the scale of symmetry
breaking~\cite{Bardeen:1978nq,Bardeen:1986yb}.} Although the original
proposal has been ruled out by experimental
data~\cite{asano81_searc_rare_decay_mode_axion}, extensions to this
model have been proposed by considering a higher scale of symmetry
breaking, causing that the mass of these axions is rather
small~\cite{kim79_weak_inter_singl_stron,shifman80_can_confin_ensur_natur_cp,dine81_simpl_solut_to_stron_cp,Zhitnitsky:1980tq}. These
generalizations belong to the so-called invisible axion models.

There exist models which predict pseudoscalar particles with similar
features as the QCD axion, and they have been collectively called \emph{axion-like particles} (ALPs).\footnote{The difference
  between models involving axions and ALPs is their number of free
  parameters: while the former is determined only by the scale of
  symmetry breaking, the latter is determined by their mass and
  characteristic energy scale as independent quantities.} Among these
models, we can mention: models with extra
dimensions~\cite{Witten:1984dg,Svrcek:2006yi,dienes00_invis_axion_large_radius_compac},
two-Higgs-doublet
models~\cite{branco12_theor_phenom_two_higgs_doubl_model,haber85_searc_super},
majorons~\cite{chikashige81_are_there_real_golds_boson,gelmini81_left_handed_neutr_mass_scale},
relaxions~\cite{Graham:2015cka,Antipin:2015jia,Gupta:2015uea},
familons~\cite{wilczek82_axion_famil_symmet_break,carone12_pseud_famil_dark_matter,jaeckel14_famil_wispy_dark_matter_candid},
gravitationally-induced
axions~\cite{Duncan:1992vz,Mielke:2006zp,Castillo-Felisola:2015ema},
etc. Although either axions and ALPs couple to the electromagnetic
Pontryagin density, a crucial difference between them arises by
considering their coupling to the gluon Pontryagin density: while
axions do couple to the latter, ALPs do not.

The characteristic Lagrangian of models involving ALPs has the
following form
\begin{equation}
  \label{ALPlag}
  \Lag_{\textsc{alp}} = \frac{1}{2}\partial_\mu a \partial^\mu a - \frac{1}{2}m_a^2 a^2 - \frac{g_{\textsc{alp}}}{4} a  F_{\mu\nu}\tilde{F}^{\mu\nu},
\end{equation}
where $m_a$ is the mass of the ALP, $g_{\textsc{alp}}$ is a
model-dependent ALP-photon coupling, $F_{\mu\nu}$ and $\tilde{F}_{\mu\nu}$ are
the electromagnetic field strength and its dual, respectively. The
interaction between ALPs and photons provides a decay channel for
\mbox{$a\to\gamma\gamma$}, which plays a key role in experimental searches. To
leading order, the decay width for this process is given
by~\cite{Beringer:1900zz,Cadamuro:2011fd}
\begin{equation}
  \label{ALPsDW}
  \Gamma\left(a\to\gamma\gamma\right) = \frac{g_{\textsc{alp}}^2 m_a^3}{64 \pi}.
\end{equation}

\begin{figure}[t] 
  \centering
  \begin{tikzpicture}[thick,scale=1.5]
    \draw[particle] (-2,-1) -- (2,-1);
    \draw[left] (-2,-1) node {$e$}; 
    \draw[right] (2,-1) node {$e$}; 
    \draw[photon] (-1,-1) -- (0,0);
    \draw[left] (-.5,-.3) node {$\gamma$};
    \draw[dashed] (-1,1) -- (0,0);
    \draw[left] (-.5,.3) node {$a$};
    \draw[photon] (0,0) -- (2,0);
    \draw[above] (1,0) node {$\gamma$};
    \fill[black] (0,0) circle (.14cm);
  \end{tikzpicture}
  \caption{Primakoff effect for ALPs, analogous to the $\pi^0$
    case. This process might be relevant for detecting solar and
    cosmological ALPs through helioscope and haloscope techniques,
    respectively, as shown in Ref.~\cite{Sikivie:1983ip}.}
  \label{fig:Primakoff}
\end{figure}
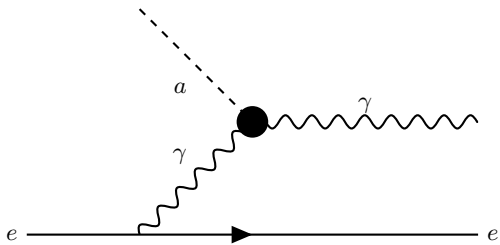

Axions and ALPs can be converted into photons (and vice-versa) through
the Primakoff effect~\cite{Halprin:1966zz}, when a strong external
electromagnetic field is present (see Fig.~\ref{fig:Primakoff}). This
process might induce an ALP-photon oscillation as well, similar to the
case of massive neutrinos~\cite{Sikivie:1983ip}. This interaction 
changes the polarization of photons traveling in external magnetic
fields, providing an additional mechanism in order to detect these
pseudoscalar particles~\cite{Maiani:1986md}. The ALP-photon conversion 
could cause an apparent dimming of distant sources as well, affecting the 
luminosity-redshift relation of Ia supernovae, the dispersion of quasar
spectra, and the spectrum of the CMB~\cite{Mirizzi:2006zy}. Constraints
on these models have been collected in Fig.~\ref{fig.Gagg}.

\section{Astrophysical Implications\label{astro_impl}}

As we aforementioned, current experimental data suggests an extragalactic
origin for UHECRs with energies above the GZK limit~\cite{Aab:2017tyv}. Here, we consider
particles beyond the stantard model, such as dark photons and ALPs, as candidates for producing 
these ultra high energy events. These particles oscillate into ordinary photons and vice-versa, allowing
them to travel across the Universe essentially without decaying. The decay length of a particle 
is given by
\begin{equation}
  L = \frac{E_{p}}{\Gamma_{p} m_{p} },
\end{equation}
where $E_{p}$, $\Gamma_{p}$, and $m_{p}$ are the energy, decay width and
mass of the particle $p$, respectively.  If we require that the decay length has 
to be at least of the order of magnitude of the observed
universe $R_U$ as considered in Ref.~\cite{gorbunov01_axion_partic_as_ultrah_energ_cosmic_rays},
one finds the following condition
\begin{equation}
  \label{cond}
  R_U \lesssim L \equiv \frac{E_{p}}{\Gamma_{p} m_{p} } \quad  \Rightarrow \quad \Gamma_{p} \lesssim \frac{E_{p}}{R_U  m_{p} }.
\end{equation}
By means of Eq.~\eqref{ALPsDW}, Eq.~\eqref{DPsDW} and Eq.~\eqref{cond}
it is possible to establish a restriction on the parameter space in
order for the ALPs and DPs to reach the Earth from distances beyond the
$R_{\textsc{gzk}}$ radius. These restrictions are:
\begin{itemize}
\item For dark photons:
  \begin{equation}
    \label{CDPs}
    g_{\textsc{dp}} \lesssim \left(  \frac{3  E_{A^\prime}}{\alpha_{\text{em}}  R_U  m_{A^\prime}^2  \Big( 1+\frac{2 m_{\ell}^2}{m_{A^\prime}^2}  \Big)       \sqrt{1-\frac{4 m_{\ell}^2}{m_{A^\prime}^2}}}  \right)^{1/2}.
  \end{equation}

\item For ALPs:
  \begin{equation}
    \label{CALPs}
    g_{\textsc{alp}} \lesssim \left(  \frac{64 \pi E_{a}}{R_U m_{a}^4}   \right)^{1/2}.
  \end{equation}
\end{itemize}

\begin{figure}[H] 
  \includegraphics[width=\linewidth]{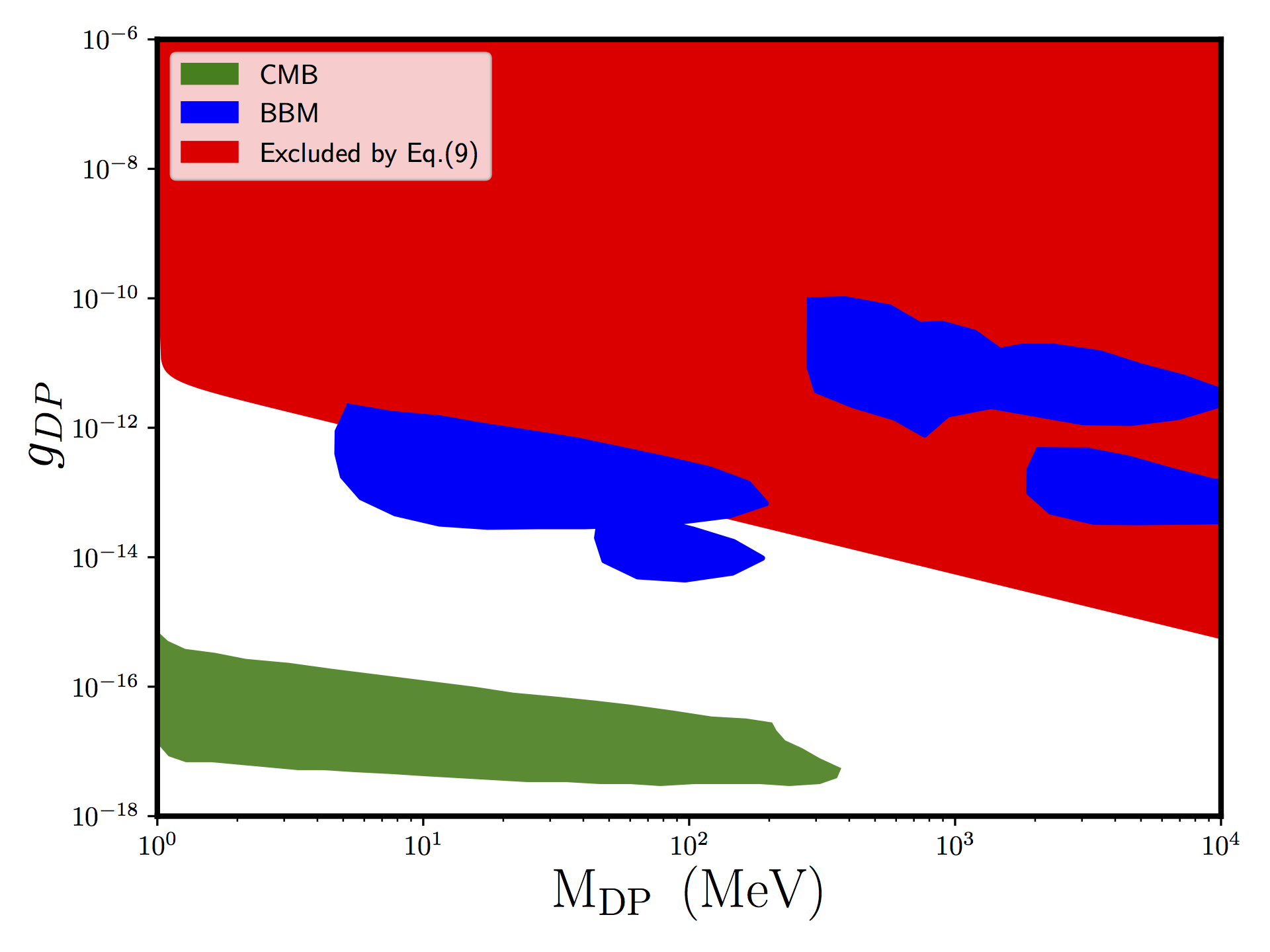}
  \caption{The region in red is excluded by Eq.~\eqref{CDPs} for \mbox{$E_{A'} \sim \SI{e20}{\eV}$},
    the regions in blue by Big Bang Nucleosynthesis~\cite{Fradette:2014sza}, and the
    region in green by the CMB~\cite{Fradette:2014sza}.}
  \label{fig.DPlim}
\end{figure}

In addition to the limits imposed by Eq.~\eqref{CDPs} for models involving dark photons, their coupling and mass are constrained by Big Bang Nucleosynthesis and the physics of the CMB, as shown in Fig.~\ref{fig.DPlim}. It is manifest that these limits do not alter the cosmological evolution of the Universe. Experimental bounds have been found by considering direct detection~\cite{DelNobile:2015uua}, colliders and fixed-target experiments~\cite{Lees:2017lec}, indirect detection~\cite{Batell:2009zp,Schuster:2009au}, and Supernovae data~\cite{Kazanas:2014mca}. These bounds are consistent with the ones found in the present analysis. Furthermore, dark photons could also be produced through dark matter annihilation at the center of the Earth and Sun, and may be detected by IceCube, or Alpha Magnetic Spectrometer (AMS-02)~\cite{Feng:2015hja,Feng:2016ijc}. As shown in Ref.~\cite{Feng:2016ijc}, the sensitivity of AMS-02 allows to search for dark photons with $m_{A'}\sim \SI{100}{\MeV}$ and \mbox{$10^{-11}\leq g_{\textsc{dp}}\leq 10^{-8}$}. This region in the parameter space is excluded in Fig.~\ref{fig.DPlim}. Thus, dark photons as candidates for UHECRs cannot be detected by AMS-02 within its present sensitivity. 

\begin{figure}[H] 
  \includegraphics[width=\linewidth]{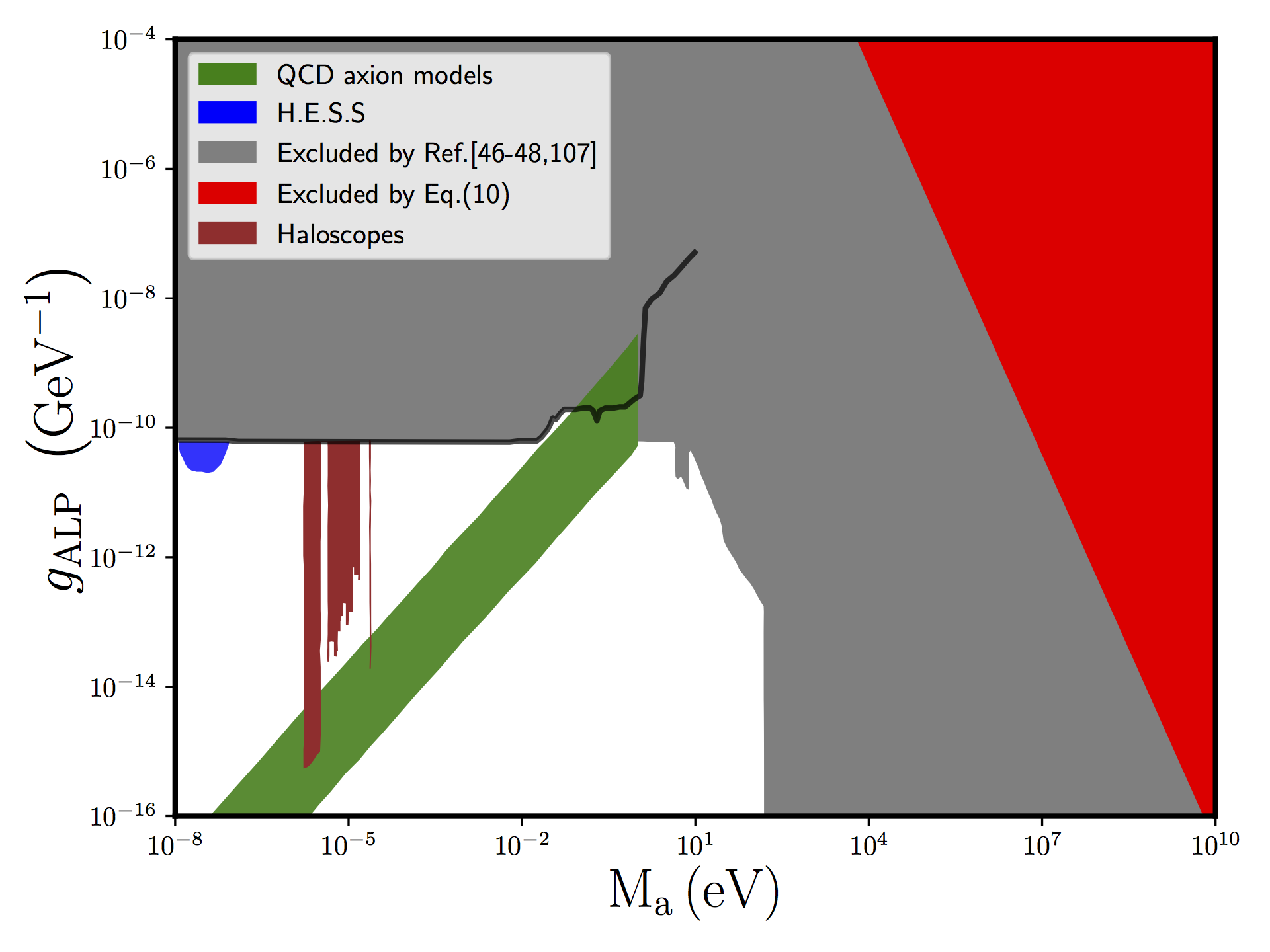}
  \caption{The region in red is excluded by Eq.~\eqref{CALPs} for \mbox{$E_a \sim \SI{e20}{\eV}$}, the region in green for typical QCD axion models~\cite{DiLuzio:2016sbl}, 
    and the regions in grey, blue and brown by Refs.~\cite{Anastassopoulos:2017ftl,Cadamuro:2011fd,DiLuzio:2016sbl,Anastassopoulos:2017kag}.}
  \label{fig.Gagg}
\end{figure}

In the case of ALPs, the limits imposed by Eq.~\eqref{CALPs} are tighter than those of Ref.~\cite{gorbunov01_axion_partic_as_ultrah_energ_cosmic_rays}, allowing to constrain a region in the parameter space that was not excluded in previous analysis~\cite{Anastassopoulos:2017ftl,Cadamuro:2011fd,DiLuzio:2016sbl}. In addition to the present exclusion limits imposed on ALPs models, their parameter space have been constrained by experimental data coming from ``light shining through a wall'' technique used in the GammeV experiment~\cite{Chou:2007zzc}, gamma rays data from H.E.S.S observations~\cite{Abramowski:2013oea}, helioscope technique used by the CAST collaboration~\cite{Anastassopoulos:2017ftl}, cosmological data~\cite{Cadamuro:2011fd}, among others, which have been collected in Fig.~\ref{fig.Gagg}. It is worth mentioning that, although the bounds presented here do not reach the sensitivity of current experiments, such as CAST, nor future ones like ALPS-II and IAXO, they still allow for detecting ALPs with lower energies by means of gamma ray telescopes as proposed in Ref.~\cite{Hooper:2007bq}.  

By considering the arclength described by the ultra high energy lepton pair produced
through the dark photon decay, to be less than the Earth diameter, i.e., $D_E\,\phi \leq 2 R_\oplus$, 
where $R_\oplus \simeq \SI{6370}{\km}$ is the radius of the Earth, we find

\begin{align}
\label{DT}
D_E \leq \frac{2 R_\oplus m_{\textsc{dm}}}{\sqrt{m_{A'}^2 - 4m_\ell^2}}\, .
\end{align}
It is worth noticing that if the DP decay occurs inside of a region with radius given by Eq.~\eqref{DT}, the two final leptons can reach the Earth and may produce two super pre-showers correlated in time and space. These signals can be tested in future cosmic rays experiments, as we will discuss in Sec.~\ref{discuss}.

\begin{figure}[H] 
  \includegraphics[width=\linewidth]{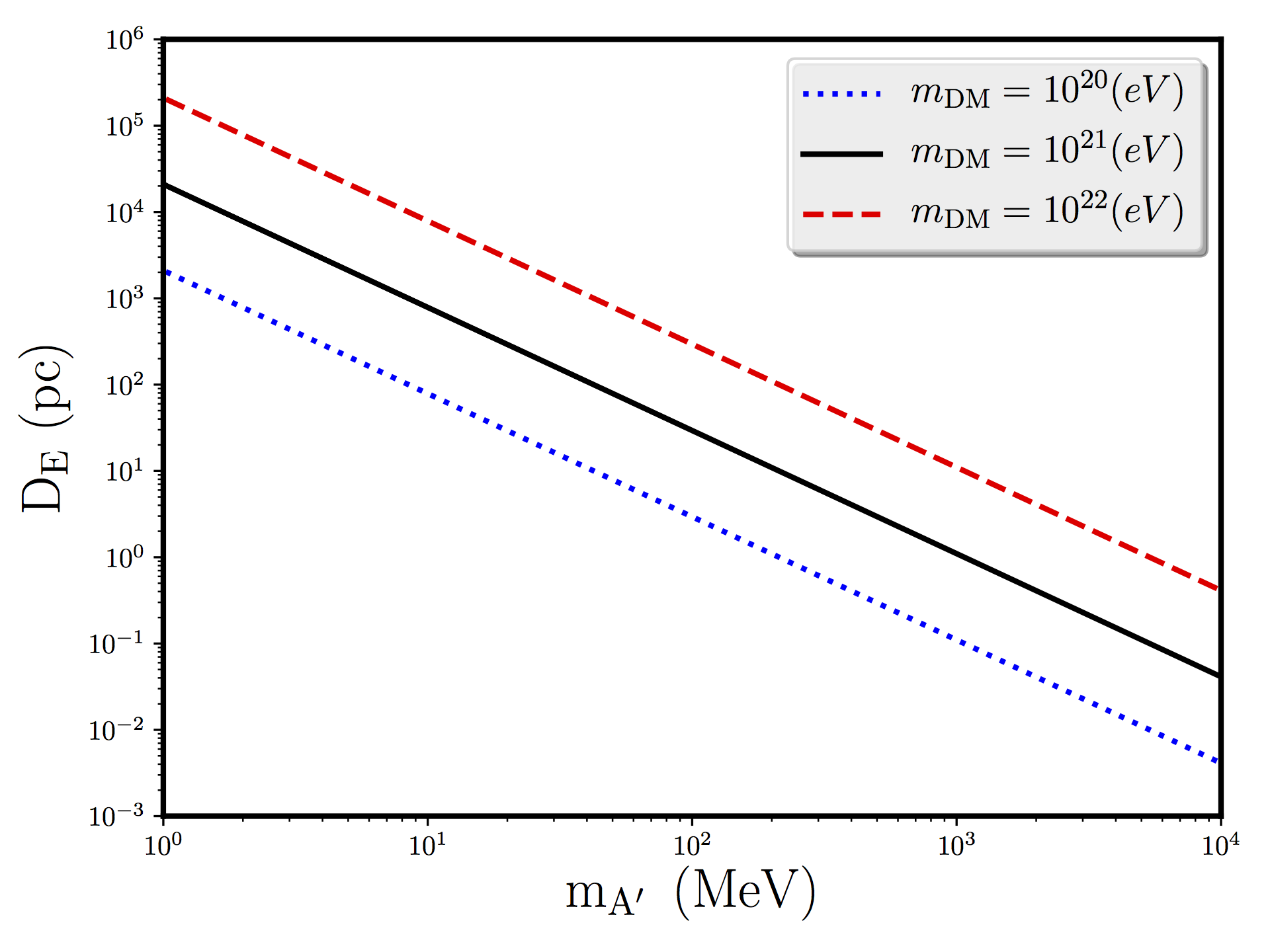}
  \caption{The regions below the lines are allowed for final lepton pairs which may produce two highly correlated UHECR events on Earth.}
  \label{fig.DE}
\end{figure}

\section{Discussion\label{discussion}}
\label{discuss}

In this paper, we have discussed scenarios where UHECRs are produced by
beyond the standard model particles, namely, dark photons and 
axion-like particles. It is known that these particles must
satisfy some conditions in order to produce the UHECRs events:
\begin{itemize}
\item long lived, in order to travel cosmological distances,
\item weakly interacting with radiation,
\item produced significantly at the source,
\end{itemize}
and, in addition, they have to interact strongly enough near
the Earth. Under these assumptions and criteria explained 
in Sec.~\ref{astro_impl}, we constrained the parameter 
space of the proposed scenarios, enhancing the present exclusion limits 
by considering these particles as responsibles for the UHECRs.

Although the constraints imposed by observations in the ALP
models are much tighter than former estimations (see
Ref.~\cite{gorbunov01_axion_partic_as_ultrah_energ_cosmic_rays}), it
can be seen in Fig.~\ref{fig.Gagg} that our constraints cannot match
the sensitivity of neither CAST 2017. However, the limits presented 
here exclude a new parameter region in which both \(m_a\) and \(g_{\textsc{alp}}\) are
larger in comparison with the previous analysis of Ref.~\cite{Cadamuro:2011fd}.

On the other hand, the assumption that UHECRs are produced by dark
photons decays, excludes a large region of the parameter space,
providing limits as satisfactory as those imposed by Big Bang
Nucleosynthesis. Furthermore, the region below $g_{\textsc{dp}}< 10^{-15}$ 
can only be constrained using data from the CMB. It is worth mentioning that 
the imposed constraints are compatible with the sensitivity of several current
experiments~\cite{Fradette:2014sza}.

It is possible to find place for improvement in the UHECR research
during the next decade, by considering the proposal of organizing the
existing professional detectors together with smart devices---such as
mobiles or tablets---as a network capable of global monitoring and
analysis of muons coming from showers produced by primary cosmic rays
in the atmosphere. A wide spatial distribution of the devices
contributing to such a network will help to detect and study possibly
correlated cosmic ray events, known as ensembles of cosmic rays
\cite{Dhital:2017yuu}. These ensembles might be composed of widely
distributed events spanning even the whole cosmic-ray energy spectrum
which might be observable only by widely spread and possibly dense
network of detectors, while even the largest individual observatories
might fail to give a trigger. The involvement of the users of smart
devices will increase the collective surface of the whole network only
mildly, but their geographical spread will significantly increase the
capability of the network to observe and analyze possibly existing
large scale correlations in the cosmic ray data, as we may expect for dark photons decaying inside the region showed in Fig.~\ref{fig.DE}. The proposal of a
global cosmic-ray network is being implemented by the Cosmic Rays
Extremely Distributed Observatory (CREDO) Collaboration
\cite{Sushchov:2017kdi,Dhital:2017yuu,Homola:2017pc}. It should be
also noted that there are already three smartphone applications
enabling the cosmic-ray detection mode (i) Distributed Electronic
Cosmic-ray Observatory (DECO) \cite{Vandenbroucke:2015ila}; (ii)
Cosmic RAYs Found In Smartphones (CRAYFIS) \cite{Whiteson:2014kca};
and (iii) CREDO Detector \cite{credo.science}. The cascade approach
proposed by CREDO will help to probe both classical and exotic
scenarios, whenever cascades of particles/photons are initiated above
the atmosphere, it will also be an instrument prepared to detect
theoretically unexpected manifestations of New Physics which can be
observed as zero-background events of correlated excesses of
cosmic-ray rates recorded by distant detectors \cite{Homola:2017pc}.

Recently, the DES collaboration has released their results from the
first year of data, including an analysis of the weak lensing mass map
with sources at redshift \(\num{.2} < z <
\num{1.3}\)~\cite{Chang:2017kmv}, providing for the first time an
astronomical survey of possible dark matter clusters. This could be useful 
to analyse the correlation between the location
of UHECR sources and the anisotropies in the dark matter distribution,
which might used to test the hypothesis that primary UHECRs are
composed by dark matter and/or that they are generated by the decay of
super-heavy dark matter.

\begin{acknowledgments}
 The authors thanks to W.~Bietenholz, Y.~Bonder and S. Troitsky for critical remarks. 
 O.C-F. would like to thank to the Physics Department of 
  \emph{Universidad de La Serena} for the hospitality during the
  (partial) development of this paper. The work of O.C-F is supported 
  by the projects PAI-79140040 and FS0821 (CONICYT--Chile), C.C. by UNAM-DGAPA-PAPIIT
  Grant IA101116 and UNAM-DGAPA postdoctoral fellowship (M\'exico), J.Z-S by the 
  Grant \emph{Becas Chile} No. 74160012, CONICYT, and P.H. by the National Science Centre (Poland) grant No. 2016/23/B/ST9/01635.
\end{acknowledgments}

\bibliography{sumbib}

\end{document}